\documentclass[prd,twocolumn,showpacs,floatfix,superscriptaddress]{revtex4}

 \pdfoutput=1

\usepackage{amsmath}
\usepackage[utf8]{inputenc}
\usepackage{graphicx}
\usepackage{epsfig}
\usepackage{bm}
\usepackage{amssymb}
\usepackage{float}
\usepackage{amsmath}
\usepackage{dcolumn}
\usepackage{cancel}
\usepackage{graphicx}
\usepackage{epsfig}
\usepackage{bm}
\usepackage[greek,english]{babel} 
\usepackage{alphabeta}
\usepackage{amssymb}
\usepackage{subfigure}
\usepackage{float}
\usepackage{amsmath}
\usepackage{dcolumn}
\usepackage{cancel}
\usepackage[colorlinks]{hyperref}
\usepackage[usenames,dvipsnames]{color}
\hypersetup{
     breaklinks=true,
    pdfstartview={FitH},    
    colorlinks=true,       
    linkcolor=blue,          
    citecolor=red,        
    filecolor=magenta,      
    urlcolor=blue,           
    anchorcolor=green,      
    linktocpage=true
}

\newcommand{\bea}{\begin{eqnarray}}
 
\newcommand{\eea}{\end{eqnarray}}
\newcommand{\bma}{\begin{pmatrix}}
\newcommand{\ema}{\end{pmatrix}}
\newcommand{\be}{\begin{equation}}
\newcommand{\ee}{\end{equation}}
\newcommand{\beno}{\begin{equation*}}
\newcommand{\eeno}{\end{equation*}}

\def\doi{http://doi.org}

\begin{document}

\title{Effect of Accretion on the evolution of Primordial Black Holes in the context of Modified Gravity Theories}

\author{Shreya Banerjee}
\email{shreya@iitism.ac.in}
\author{Aritrya Paul}
\email{22ms0028@iitism.ac.in}

\affiliation{Department of Physics,
Indian Institute of Technology (Indian School of Mines),
Dhanbad, Jharkhand-826004, INDIA}

\begin{abstract} 
We investigates the effect of accretion of cosmic fluid
on the evolution of Primordial Black Holes (PBHs) within the framework of
Modified gravity theories. We consider a general form of the Hubble parameter, reflecting a general class of modified gravity theories and bouncing models. We then study the effect of such modified dynamics on PBH in the presence of Hawking radiation and accretion of surrounding materials. We investigate how the evolution of PBHs is influenced by accretion across different cosmological eras, considering the radiation, matter, and dark energy-dominated phases like phantom and quintessence for linear equation of state.  We further incorporated Non-linear Equations of State such as Chaplygin Gas, Modified Chaplygin
gas, Van der Waals model, Polytropic Fluid model. The study systematically
analyzes the mass variation of PBHs in the presence of such different cosmological environments. The results will contribute
to the understanding of PBH formation and evolution in modified theory of
gravity, and their possibility of being detected with future experiments.
\end{abstract}


\maketitle

\section{Introduction}
Observational physics is currently in its golden age. Wealth of new observational results
both at cosmological and astrophysical scales are being uncovered. The temperature variations of
the CMBR bear testimony of minute fluctuations in the density of the primordial universe
which have been measured and verified by experiments like WMAP and PLANCK \cite{2013ApJS..208...19H, Planck:2018vyg}.
 Experiments build to probe at galactic scales have mapped the
distribution of matter and have found the presence of unknown/unexplained stuff (hence
named it as dark matter) in most parts of the galaxies. There are a lot of differing theories about what dark
matter is made of, and how we might actually find it. The $\Lambda$CDM model is the current `standard model' of Big Bang
cosmology. Consensus is that it is the simplest model that can account for the various
measurements and observations relevant to cosmology. However, as mentioned, every new theory has its own set of drawbacks and fundamental issues which calls for further introspection. This model assumes: GR is the correct theory of gravity on
cosmological scales. The current standard model of cosmology
($\Lambda$CDM+inflation) however faces some serious fundamental drawbacks, unexplained issues till date. This has intrigued scientists to probe for alternate theories of gravity, go beyond Einstein's GR.

Modified gravity forms a very broad class of theories that aim to overcome the existing problems in $\Lambda$CDM model and GR. For example, 
the non-renormalizability of GR, the cosmological constant problem, tension with cosmological data, and so on. Such theories lead to
improved cosmological evolution, both at the background as well as at the perturbation levels
\cite{CANTATA:2021ktz, Capozziello:2011et}. Such gravitational modifications can be constructed by modifying Einstein-Hilbert
action of GR, resulting to theories like $f(R)$ gravity
\cite{DeFelice:2010aj, Nojiri:2010wj, Starobinsky:2007hu, Cognola:2007zu, Zhang:2005vt, Papanikolaou:2021uhe}, Gauss-Bonnet and $f(G)$ gravity \cite{Nojiri:2005jg,DeFelice:2008wz,Zhao:2012vta}, Lovelock gravity \cite{Lovelock:1971yv,Deruelle:1989fj}, $f(T)$ gravity \cite{Cai:2015emx,Chen:2010va,Bengochea:2010sg,Benetti:2020hxp}, $f(T, T_G)$ gravity \cite{Kofinas:2014owa,Kofinas:2014daa}, etc.
 Most of these theories lead to modification of the evolution of the Hubble parameter from that of GR at the background level. 

Primordial black holes (PBHs) are considered to form in the very early universe
out of the gravitational collapse of very high overdensity regions, whose energy density is higher than a critical threshold \cite{Khlopov:2008qy,Carr:2020gox}. They were introduced back in ‘70s \cite{Zeldovich:1967lct,1974MNRAS.168..399C,1975ApJ...201....1C}.
 Remarkably, as per recent arguments, PBHs can account for a part
or the totality of the dark matter density \cite{Chapline:1975ojl,Belotsky:2014kca}, provide an explanation for the LSS formation through
Poisson fluctuations \cite{Meszaros:1975ef,Afshordi:2003zb}, they can also seed the supermassive black holes
residing in galactic centres \cite{Bean:2002kx,1984MNRAS.206..315C}. Furthermore, PBHs are associated as well
with numerous gravitational-wave (GW) signals originated from both binary merging
events and stochastic cosmology \cite{Sasaki:2018dmp,LISACosmologyWorkingGroup:2023njw}. They have been extensively analysed in the context of modified gravity theories \cite{Kawai:2021edk,Yi:2022anu,Zhang:2021rqs} and bouncing cosmology \cite{Carr:2011hv,Carr:2014eya,Quintin:2016qro,Chen:2016kjx,Banerjee:2022xft,Papanikolaou:2024fzf}. Significant modifications to the perturbation power spectra have been found due to the presence of modified background dynamics.

 Long time ago Hawking showed that black holes can emit thermal radiation through relativistic quantum effects, leading to black hole evaporation \cite{Hawking:1975vcx}. It was also found that small mass PBHs may have evaporated through the Hawking radiation. The lifetime of such PBHs are comparatively long, comparable to the age of the universe, owing to their small masses.  Thus, the existence of such PBHs may become possible through the detection of the emitted Hawking radiation.

PBHs can also accrete matter from the surrounding medium which may lead to an increase in the lifetime of these black holes. Such long surviving PBHs may constitute a major portion of dark matter \cite{BLAIS200211,Blais:2002gw,Barrau:2003xp}.
Recent studies have examined the impact of phantom energy accretion on the evolution of primordial black holes (PBHs) \cite{jamil2011primordial}. These studies indicate that the mass of PBHs decreases over redshift, thus increasing over time. Additionally, the effects of matter, radiation, and dark energy (both phantom and quintessence types) accretion have been explored within the framework of modified $f(T)$ gravity theory. The evolution of PBH mass due to the accretion of modified Chaplygin gas has also been investigated in the context of $f(T)$ gravity theory\cite{debnath2015evolution}.

In this paper, we investigate how the evolution of PBHs is influenced by accretion across different cosmological eras, considering the radiation, matter, and dark energy-dominated phases. We perform the analysis without considering any particular model but through a specific parametrization of the Hubble parameter that reflects the effect of modified background dynamics. By examining a specific form of the Hubble parameter that represent a variety of modified gravity and bouncing models and incorporating nonlinear equations of states (EOS)\cite{sadhukhan2023non}, we aim to estimate the mass evolution of PBH by varying cosmological environments. Here we have considered Chaplygin gas model, Modified chaplygin gas model, Van-der-waals fluid and polytropic model for the non-linear cases.

This paper has been organised as follows. In Section \ref{Overview}, we give a brief overview of PBHs and the different processes through which the mass of PBH can change. In Section \ref{PBH Evol.}, we discuss the general methodology that we have undertaken and evaluated the general expression for the rate of change of mass of PBH. We then analyse the rate of change of mass in the presence of different cosmological fluids/eras. We then extended our analysis to the case of non-linear models like Chaplygin gas, Vander Waals, Polytropic fluid, etc. Finally we conclude in Section \ref{Conclusions}.

\section{Overview}
\label{Overview}
\noindent
In this section we give a brief overview of Primordial Black Hole. It is known that Black Holes are the final stage of evolution of a collapsing massive star whose mass exceeds twice the mass of the sun. There are another kind of Black Holes which are formed in early stage of the universe due to inhomogeneities in the scalar field\cite{feinstein1995scalar}. Primordial Black Holes (PBHs) are hypothetical black holes that are believed to have formed soon after the Big Bang. In the early radiation-dominated universe, regions of critical density containing subatomic matter could have gravitationally collapsed, resulting in the creation of PBHs without the need for a supernova, which is the process that forms black holes today. The masses of these primordial black holes are much smaller than those formed by the collapse of massive stars. PBHs offer a unique window into the early universe, providing valuable insights into its extreme conditions and cosmological evolution. Accretion, the process by which PBHs accumulate mass, plays a crucial role in their evolution and can significantly influence their properties over cosmic time.\cite{abramowicz2013foundations}\cite{lasota2016black}. Primordial black holes (PBHs) can increase their mass through gravitational attraction, accreting material from their surroundings. This accretion process is especially notable during the early stages of PBH formation and in dense regions of the Universe where matter is more concentrated. Additionally, the rate at which PBHs accrete mass depends on the density and composition of the surrounding environment. The mass of PBHs can decrease due to Hawking radiation\cite{page2005hawking}\cite{auffinger2023primordial}. Hawking radiation, a theoretical prediction by physicist Stephen Hawking, significantly influences the evolution of primordial black holes (PBHs) by causing their gradual mass reduction over time. According to Hawking's theory, particles continuously escape from the black hole. This quantum-level process near the black hole's boundary leads to the emission of radiation, known as Hawking radiation.
In this paper, we investigate how the evolution of PBHs is influenced by accretion of cosmic fluids. The rate of change of mass of PBH depends highly on the nature of background dynamics, hence on the nature of evolution of the Hubble parameter. For a given general form of the Hubble parameter, we estimate the mass evolution of PBH for both linear and non-linear of equation of state. Here we have considered Chaplygin gas model, Modified chaplygin gas model, Van-der-waals fluid and polytropic model for the non-linear cases. The observation of supernovae has led to the prediction that the universe is currently in a phase of accelerated expansion.\cite{riess1998observational}\cite{perlmutter2003measuring}. Understanding the fundamentals behind the current cosmic acceleration has been a constant motivation for scientists. Various efforts have been made within the standard cosmological model to explain the universe's expansion \cite{sahni2000case, padmanabhan2003cosmological, peebles2003cosmological}. The expansion of the universe can be understood by considering that the background of the universe is predominantly influenced by dark energy (DE), which currently plays a major role. Observations indicate that dark energy accounts for approximately 70\% of the total energy density of the universe, which has negative pressure\cite{akarsu2014cosmology}. In this work, we have considered various models of dark energy namely, quintessence$(-1 < w <-1/3)$\cite{Chiba:1999ka}\cite{Amendola:1999er}, phantom phase$(w < -1)$\cite{caldwell2003phantom} etc. We have also considered the late time universe when the universe was governed by radiation\cite{lora2013pbh} and then matter era\cite{FREESE20021}. We have predominately investigated the effect of their presence on the net mass change of PBHs with cosmic evolution. In the following sections, we discuss our methodologies and results in this regard.

\section{PBH Evolution for a general Hubble Parameter}
\label{PBH Evol.}

There are two primary ways in which low mass PBH can loose or gain mass.
\noindent
In low mass PBH, mass loss occurs through evaporation due to Hawking Radiation. The rate of mass loss for PBH is given by \cite{PhysRevD.13.198}
\begin{equation}\label{1}
\dot{M}_{eva}=-\frac{a_{H}}{256G^2\pi^3M(t)^2}
\end{equation}
where $a_H$ is the black body constant.

Now if we consider the presence of cosmic fluid in and around the PBH, this will lead to an increase in mass of the PBHs through accretion expressed as \cite{babichev2004black},

\begin{equation}\label{2}
\dot{M}_{accr}=4\pi AM(t)^2(\rho + p)
\end{equation}
where $A$ is the accretion efficiency and $\rho$ is the density of the cosmic fluid and $p$ is the pressure. We at first consider linear relation between density and pressure given by $p=w\rho$, where $w$ is the corresponding equation of state. For radiation dominated era $(w = 1/3)$, matter era $w = 0$, dark energy era $w<-1/3$ and phantom accretion $w < -1$.\\
So, the resulting PBH mass evolution is given by,
$$\dot{M}=\dot{M}_{eva}+\dot{M}_{accr}$$
\begin{equation}\label{3}
\dot{M}=-\frac{a_{H}}{256G^2\pi^3M(t)^2}+4\pi AM(t)^2(\rho + P)
\end{equation}
Using $a=a_0/(1+z)$ and choosing the present value of the scale factor $a_0=1$, we have
\begin{equation}\label{4}
\frac{dz}{dt}=-H(t)(1+z)
\end{equation}
Without loss of generality, we consider the following parametrization of the Hubble parameter
\begin{equation}\label{5}
   H(t)=\alpha t^{\beta} 
\end{equation} 
where the values of the constant $\alpha$ and the power $\beta$ will determine the model. They constitute our main parameters which will be constrained. This parametrization not only apply to a class of modified gravity models \cite{cai2016f, yang2011new, harko2011f, fernandes20224d, xu2019f,Odintsov:2018nch} but also to a variety of bouncing models \cite{Quintin:2014oea,Cai:2007qw,Quintin:2015rta,Cai:2012va,Singh:2022jue,Chen:2022usd}. The parameter $\beta$ can be either integer or fraction, both positive and negative signs. Through the following analysis, we consider all such possible cases/forms of the Hubble parameter and see their effect on the evolution of PBH. \\
Eq. (\ref{4}) will now be,
\begin{equation}\label{6}
\frac{dz}{dt}=-\alpha t^{\beta}(1+z)
\end{equation}
The mass evolution for PBH in terms of redshift parameter is now given by,
\begin{equation}\label{7}
\frac{dM}{dz}=-\frac{dM}{dt}\frac{1}{\alpha t^{\beta}(1+z)}
\end{equation}
By using Eq. (\ref{3}), we can reconstruct the above equation as,
\begin{equation}\label{8}
\frac{dM}{dz}=-\frac{[-\frac{a_{H}}{256G^2\pi^3M^2}+4\pi AM^2(\rho + P)]}{\alpha t^{\beta}(1+z)}
\end{equation}

\noindent
Now, integrating Eq. (\ref{6}), we get a relation between $t$ and $z$ as,
\begin{equation}\label{9}
t=\left( \frac{\beta +1}{\alpha}ln(1+z)\right)^{\frac{1}{\beta +1}}
\end{equation}
\noindent
Substituting Eq. (\ref{9}) into Eq. (\ref{8}), we will get,
\begin{equation}\label{10}
\frac{dM}{dz}=-\frac{[-\frac{a_{H}}{256G^2\pi^3M^2}+4\pi AM^2(\rho + P)]}{\alpha (1+z)[\frac{\beta +1}{\alpha}ln(1+z)]^{\frac{\beta}{\beta +1}}}
\end{equation}
In the following section we study the behaviour of $M$ with redshift in the presence of different cosmic fluids.\\

\subsection{Different Cosmological Era}
The evolution of the universe is marked by distinct cosmological eras characterized by the dominant form of energy. Initially, the `Radiation Era' prevailed, where radiation (photons and neutrinos) dominated the energy density of the universe. This was followed by the `Matter Era', where matter (both dark matter and baryonic matter) took over as the primary energy component, leading to the formation of galaxies, stars, and other cosmic structures. Currently, the universe is in the `Dark Energy Era', where dark energy, a mysterious force driving the accelerated expansion of the universe, dominates the energy density.
\subsubsection{Radiation era }

We now consider the evolution of PBHs in the cosmological background governed by radiation. We assume that the PBH density is low enough to
ensure radiation domination. For a PBH immersed in the radiation field, the accretion of radiation leads to the increase of its mass. When a PBH accretes radiation ($w = 1/3$), the equation governing this accretion, for a linear equation of state, is given by,
\begin{equation}\label{11}
\dot{M}_{accr}=\frac{16}{3}\pi AM(t)^{2}\rho
\end{equation}
and the corresponding PBH evolution is given by,
\begin{equation}\label{12}
\dot{M}=-\frac{a_{H}}{256G^2\pi^3M(t)^2}+\frac{16}{3}\pi AM(t)^{2}\rho
\end{equation}
 We know for radiation era, $\rho=\rho_{0}(1+z)^4$, where $\rho_{0}=\Omega_{r}^{0}\rho_{c}$. $\Omega_r^0$ is the present value of the radiation density parameter and $\rho_c$ is the critical density. Therefore, from Eq.(\ref{10}), 
the change of mass $M$ of PBH with respect to $z$ for general form of Hubble parameter is given by,
\begin{equation}\label{13}
\frac{dM}{dz}=-\frac{[-\frac{a_{H}}{256G^2\pi^3M^2}+\frac{16}{3}\pi AM^2\rho_{0}(1+z)^4]}{\alpha (1+z)\left[\frac{\beta +1}{\alpha}ln(1+z)\right]^{\frac{\beta}{\beta +1}}}
\end{equation}
In all the analyses below, we try to analyse the effect of model parameter $\beta$ for a given value of $\alpha$.

\begin{figure}
\centering
    \hspace{-2mm}
    \subfigure[]{\includegraphics[width=0.88\linewidth]{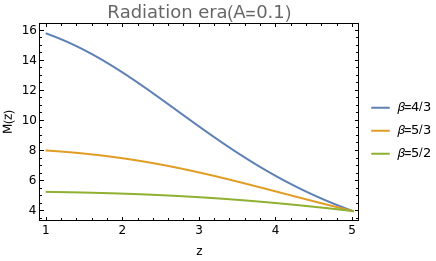}}\\
    \subfigure[]{\includegraphics[width=0.88\linewidth]{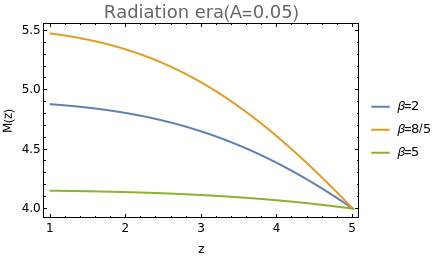}}\\
    \subfigure[]{\includegraphics[width=0.88\linewidth]{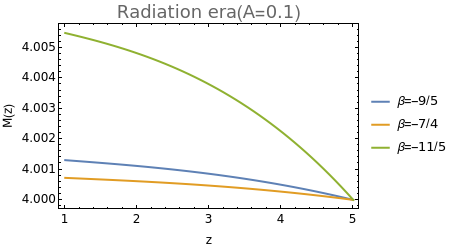}}\\
    \hspace{-1mm}
    \subfigure[]{\includegraphics[width=0.88\linewidth]{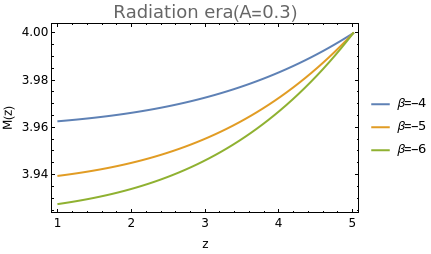}}
\caption{\it{Plot of $M(z)$ vs $z$ for radiation era for different values of $\beta$ and accretion efficiency $A$. For all the plots, the value of $\alpha=0.01$ }}
\label{rad}
\end{figure}

In Fig. \ref{rad}, we have plotted the variation of $M(z)$ with $z$ for different values of the model parameter $\beta$. We have also shown plots for different values of the accretion efficiency. As we see for positive powers of $\beta$, $M(z)$ decreases with $z$, hence increases with time. Therefore, accreting radiation leads to an increase in the effective mass of the PBH. The rate of increase depends on the value of the model parameter $\beta$. Smaller the value of $\beta$, larger is the increase in mass of PBH, for a given initial mass. However background dynamics with negative powers of $\beta$, we see in the last two plots of Fig. \ref{rad}, the PBH mass either increase very slowly with time or decrease with increase in time. Thus, we see that for background dynamics giving rise to negative power dependence of $H$ on $t$, the accretion rate is either negligible or may not result in any effective increase in the mass of PBH. In other words, the Hawking radiation either nearly balances accretion or overpowers it.

\noindent
\subsubsection{Matter era }

In matter dominated era $(w = 0)$, PBH mass can increase by absorbing surrounding matter at a rate for linear equation of state,
\begin{equation}\label{14}
\dot{M}_{accr}=4\pi AM(t)^{2}\rho
\end{equation}
So, PBH evolution is given by,
\begin{equation}\label{15}
\dot{M}=-\frac{a_{H}}{256G^2\pi^3M(t)^2}+4\pi AM(t)^{2}\rho
\end{equation}
For matter dominated era, we know, $\rho=\rho_{0}(1+z)^3$, where $\rho_{0}=\Omega_{m}^{0}\rho_{c}$ where $\Omega_m^0$ is the present matter density parameter. Therefore, using Eq.(\ref{10}), the change of mass $M$ with respect to $z$ for general form of Hubble parameter is now given by,
\begin{equation}\label{16}
\frac{dM}{dz}=-\frac{[-\frac{a_{H}}{256G^2\pi^3M^2}+4\pi AM^2\rho_{0}(1+z)^3]}{\alpha (1+z)[\frac{\beta +1}{\alpha}ln(1+z)]^{\frac{\beta}{\beta +1}}}
\end{equation}
\\
\begin{figure}\centering
    \subfigure[]{\includegraphics[width=0.88\linewidth]{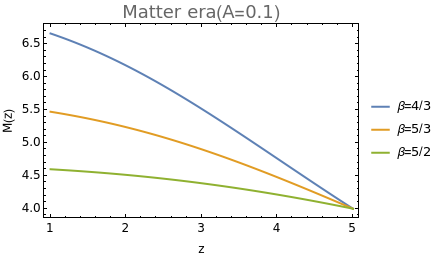}}\\
    \subfigure[]{\includegraphics[width=0.88\linewidth]{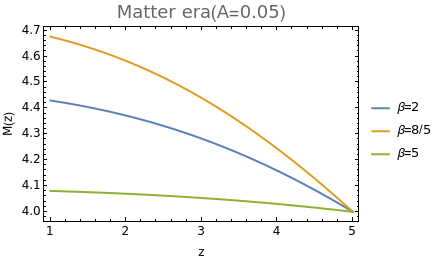}}\\
    \subfigure[]{\includegraphics[width=0.88\linewidth]{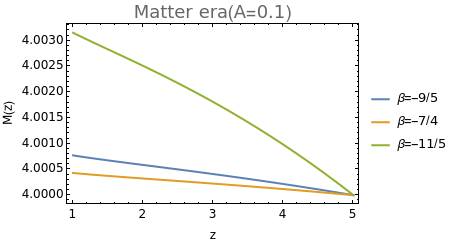}}\\
    \subfigure[]{\includegraphics[width=0.88\linewidth]{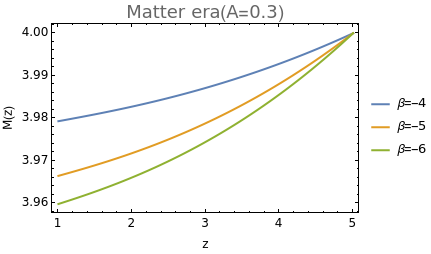}}
\caption{\it{Plot of $M(z)$ vs $z$ for matter era for different values of $\beta$ and accretion efficiency $A$. For all the plots, the value of $\alpha=0.01$ }}
\label{mat}
\end{figure}
Similar to the radiation case, in Fig. \ref{mat}, we have plotted the variation of $M(z)$ with $z$ for different values of the model parameter $\beta$ (similar to the ones we used for radiation era). We have also shown plots for different values of the accretion efficiency. Once again we see for positive powers of $\beta$, $M(z)$ decreases with $z$, hence increases with time. Therefore, accretion of matter also leads to an increase in the effective mass of the PBH. However, similar to the behaviour in the presence of radiation, here also  we see that for background dynamics giving rise to negative power dependence of $H$ on $t$, the accretion rate is either negligible or may not result in any effective increase in the mass of PBH.
\subsubsection{Dark energy Phase}
In dark energy dominated era, PBH mass changes due to dark energy at a rate,
\begin{equation}\label{17}
\dot{M}_{accr}=4\pi AM(t)^2(1+w)\rho_{DE}
\end{equation}
with 
\begin{equation}\label{18}
\rho = \rho_{0}\left(\frac{1}{1+z}\right)^{-3(1+w)} 
\end{equation}

Here we consider two different phases, (a) phantom  phase ($w<-1$) (b) quintessence phase ($-1 < w < -1/3$).

So, from Eq.(\ref{10}) the change of mass $M$ with respect to $z$ for general form of Hubble parameter is given by,\\
\begin{equation}\label{19}
\frac{dM}{dz}=-\frac{[-\frac{a_{H}}{256G^2\pi^3M^2}+4\pi AM^2\rho_{0}(\frac{1}{1+z})^{-3(1+w)}]}{\alpha (1+z)\left[\frac{\beta +1}{\alpha}ln(1+z)\right]^{\frac{\beta}{\beta +1}}}
\end{equation}
\begin{figure}
\centering
    \subfigure[]{\includegraphics[width=0.88\linewidth]{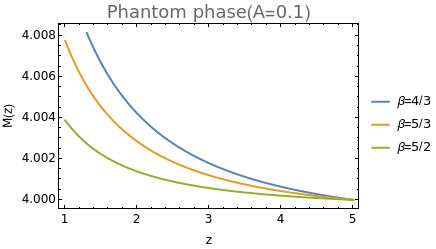}}
    \hspace{-2mm}
    \subfigure[]{\includegraphics[width=0.88\linewidth]{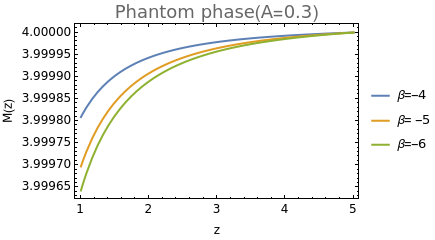}}\\
    \hspace{4mm}
    \subfigure[]{\includegraphics[width=0.88\linewidth]{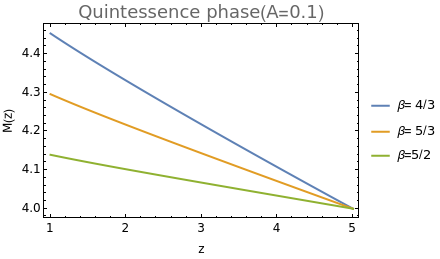}}\\
    \hspace{2mm}
    \subfigure[]{\includegraphics[width=0.88\linewidth]{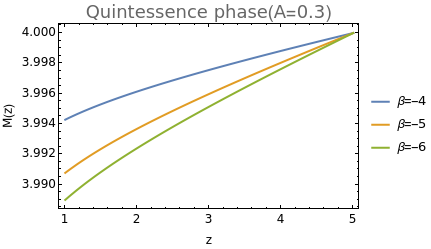}}
\caption{\it{Plot of $M(z)$ vs $z$ for dark energy era for both phantom $(w=-1.5)$ and quintessence $(w=-0.5)$ phases for different values of $\beta$ and accretion efficiency $A$.  For all the plots, the value of $\alpha=0.01$ }}
\label{dark}
\end{figure}
In Fig. \ref{dark}, we have plotted the variation of $M(z)$ with $z$ for both the quintessence and phantom phases for $w=-0.5, -1.5$ respectively. We plotted the graph for different values of the model parameter $\beta$ and accretion efficiency $A$. We see that when the dark energy behaves like a phantom fluid, the change in mass of the PBH remains either nearly constant (for positive $\beta$) or decreases with time (for negative $\beta$). This can be explained due to the negative pressure provided by the presence of phantom phase that aids the mass loss due to Hawking evaporation. 

In the presence of quintessence phase, the behaviour is similar to that of in the presence of matter or radiation, i.e. the PBH mass increases with time (hence decreases with redshift) for positive powers dependency of the Hubble parameter on time but it can tend to decrease even in the presence of dark energy fluids in the cases when the Hubble parameter has negative power dependency on time. 
\subsubsection{Presence of two or more cosmic fluids}

We will now consider the presence of more than one cosmic fluid like radiation, matter and dark energy, then the PBH mass evolution in general form is given by,
\begin{equation}\label{20}
\frac{dM}{dz}=-\frac{[-\frac{a_{H}}{256G^2\pi^3M^2}+4\pi AM^2\sum_{i=1}^{3}(1+w_{i})\rho_{i}]}{\alpha (1+z)[\frac{\beta +1}{\alpha}ln(1+z)]^{\frac{\beta}{\beta +1}}}
\end{equation}
where $i$ corresponds to the different cosmological fluid, $w_{i}$ is the corresponding equation of state and $\rho_{i}$ corresponds to the density of fluid.

\begin{figure}
    \subfigure[]{\includegraphics[width=0.88\linewidth]{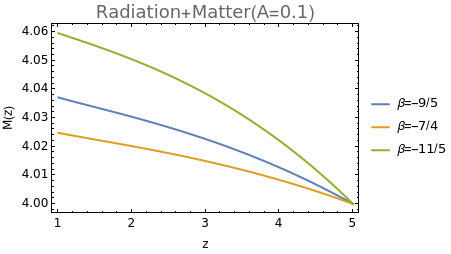}}\\
    \hspace{-4mm}
    \subfigure[]{\includegraphics[width=0.88\linewidth]{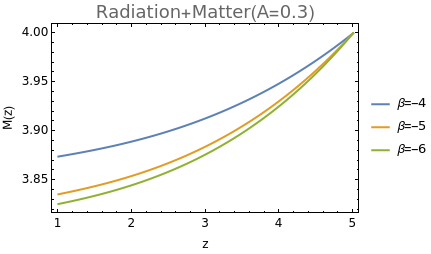}}\\
    \subfigure[]{\includegraphics[width=0.88\linewidth]{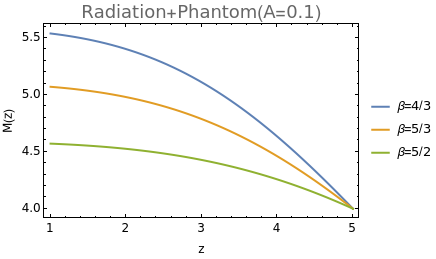}}\\
    \hspace{-3mm}
    \subfigure[]{\includegraphics[width=0.88\linewidth]{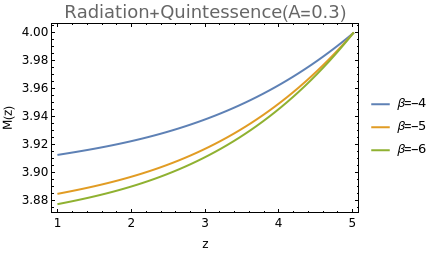}}\\
    
\caption{\it{Plot of $M(z)$ vs $z$ for the presence of two cosmic fluids for different values of $\beta$ and accretion efficiency $A$.  For all the plots, the value of $\alpha=0.1$ }}
\label{twofluid1}
\end{figure}

\begin{figure}
\subfigure[]{\includegraphics[width=0.88\linewidth]{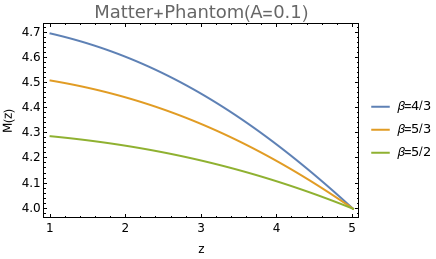}}\\
    \hspace{2mm}
    \subfigure[]{\includegraphics[width=0.88\linewidth]{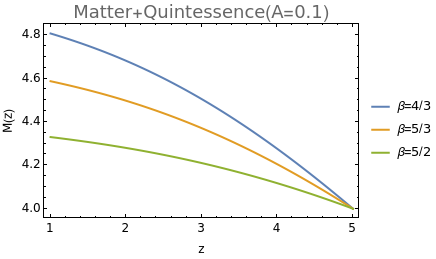}}
\caption{\it{Plot of $M(z)$ vs $z$ for the presence of two cosmic fluids for different values of $\beta$ and accretion efficiency $A$.  For all the plots, the value of $\alpha=0.1$ }}
\label{twofluid2}
\end{figure}

\begin{figure}\centering
    \subfigure[]{\includegraphics[width=0.88\linewidth]{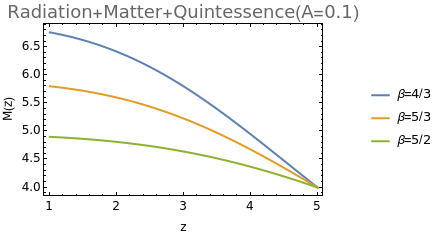}}\\
    \subfigure[]{\includegraphics[width=0.89\linewidth]{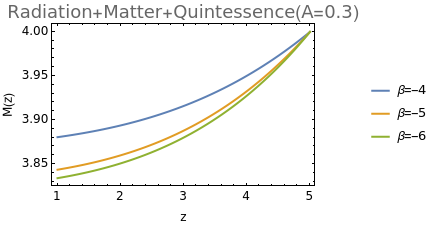}}\\
    \hspace{-5mm}
    \subfigure[]{\includegraphics[width=0.78\linewidth]{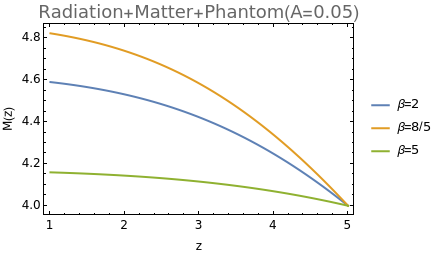}}
\caption{\it{Plot of $M(z)$ vs $z$ for the presence of two cosmic fluids for different values of $\beta$ and accretion efficiency $A$.  For all the plots, the value of $\alpha=0.1$ }}
\label{threefluid}
\end{figure}

In Fig. \ref{twofluid1}, Fig. \ref{twofluid2} and Fig. \ref{threefluid}, we have shown the accretion plots in the presence of two and three cosmic fluids in different combinations for certain values of the model parameters. As we can see, the nature of the plots remain same as before, though the rate of accretion might be more in some cases. Like before, the negative values of $\beta$ does not lead to an increase in the mass of PBH with time.
Interestingly we see when we consider the presence of either radiation or matter alongn with phantom fluid, for positive values of $\beta$, the change in mass with time no longer remains nearly constant (unlike Fig. 3a) but increases significantly (smaller the value of $\beta$, larger is the effect), thereby showing that the effect of radiation or matter dominates the behaviour of mass accretion onto PBH.


\subsection{Non-Linear Equation of State}
Till now our analysis was mainly based on the assumption of linear dependency between pressure and density. In this section we extend our analysis to the presence of different types of cosmic fluid with non-linear equation of state\cite{sadhukhan2023non}.

We know that the general expression for the rate of change of PBH mass w.r.t. the redshift is given by Eq. \eqref{10}.
These new models can be used to reconstruct different cosmological non-linear equation of state like chaplygin gas\cite{gorini2005chaplygin}, modified chaplygin gas\cite{benaoum2012modified}, Van-der waals fluid\cite{elizalde2017cosmological}\cite{jantsch2016van} and polytropic fluid\cite{Chavanis:2012uq} model with some parameter analysis. Below we discuss such models by considering different non-linear relations between the density and pressure of the cosmic fluid.
\subsubsection{Chaplygin Gas Model}

The Chaplygin gas model, originally proposed in cosmology to address the problem of dark energy, has also been employed in the study of primordial black hole (PBH) formation and evolution. Unlike the conventional perfect fluids, the Chaplygin gas model gives a unique equation of state that allows for negative pressure at high densities. This property is particularly relevant in the early Universe, where extreme conditions prevail. In the context of PBH formation, the Chaplygin gas model plays a significant role in understanding the evolution of PBH mass. The presence of negative pressure can lead to the amplification of density perturbations, which, under certain conditions, can undergo gravitational collapse to form PBHs. By incorporating the effects of the Chaplygin gas equation of state, we can investigate the conditions under which PBHs form,  and subsequent evolution.\\
In the present case, we study the model of a universe filled with the so called Chaplygin gas, which is a perfect fluid characterized by the following equation of state:
\begin{equation}\label{21}
P=-\frac{C}{\rho}
\end{equation}
where $C$ is a positive constant.\\ 
We aim to discuss the phantom phase with parametric analysis in this model. This model depends on a constant $C$. For that we consider $C > 0.$
From Eq.(\ref{10}), for chaplygin gas model, the evolution of PBH with respect to redshift parameter z is given by,
\begin{equation}\label{22}
\begin{split}
\frac{dM}{dz}=-\frac{1}{\alpha (1+z)\left[\frac{\beta +1}{\alpha}ln(1+z)\right]^{\frac{\beta}{\beta +1}}}\left[-\frac{a_{H}}{256\pi^{3}G^{2}M^{2}} \right.
\\
\left.
+4\pi AM^{2}\left[ \rho_{0}(\frac{1}{1+z})^{-3(1+w)} - \frac{C}{\rho_{0}(\frac{1}{1+z})^{-3(1+w)}}\right] \right]
\end{split}
\end{equation}
where we have also used Eq. \eqref{18}.
\begin{figure}\centering
    \subfigure[]{\includegraphics[width=0.88\linewidth]{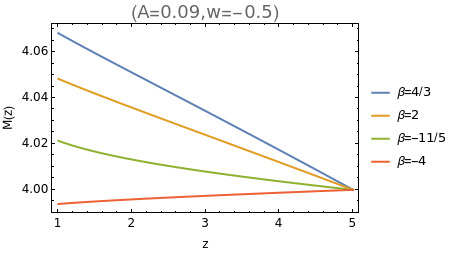}}\\
    
    \subfigure[]{\includegraphics[width=0.88\linewidth]{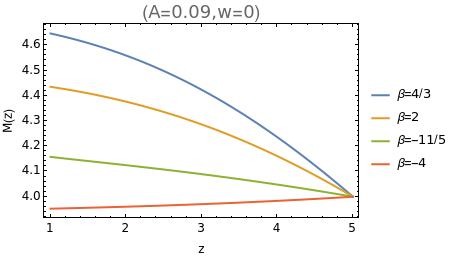}}\\
    \subfigure[]{\includegraphics[width=0.88\linewidth]{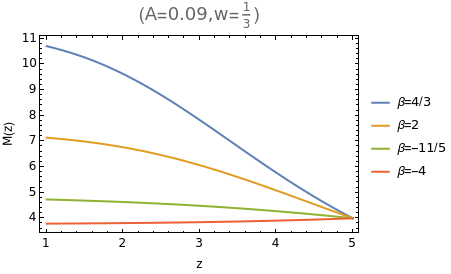}}\\
    \subfigure[]{\includegraphics[width=0.88\linewidth]{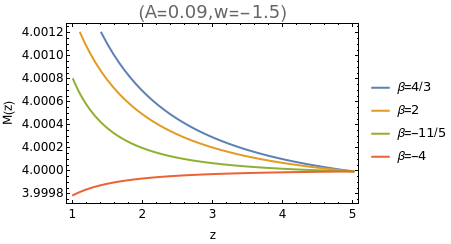}}
\caption{\it{Plot of $M(z)$ vs $z$ for Chaplygin gas model for different values of $\beta$, $w$ while accretion efficiency $A$ remains fixed. For all the plots, the value of $\alpha=0.5$, $C=10^{-8}$.}}
\label{chap1}
\end{figure}

\begin{figure}
\subfigure[]{\includegraphics[width=0.88\linewidth]{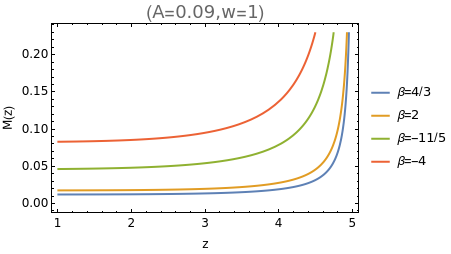}}
\caption{\it{Plot of $M(z)$ vs $z$ for Chaplygin gas model for different values of $\beta$, $w$ while accretion efficiency $A$ remains fixed. For this plot, the value of $\alpha=0.5$, $C=10^{-5}$.}}
\label{chap2}
\end{figure}

In Fig. \ref{chap1} and Fig. \ref{chap2} we have plotted the variation of $M$ with redshift for different values of $\beta$ for a given accretion efficiency. We have shown how the effective mass change of PBH changes with different values of model parameter. We have further plotted separate graphs for $w=-0.5, 0, 1/3,1,-1.5$. As we can see, irrespective of the value of $w$, for certain negative values of $\beta$ (as we have see in all the earlier cases), the PBH mass decreases with time even in the presence of Chaplygin gas like cosmic fluid. While for $w=0,1/3$, we can get a significant increase in the mass of PBH with time for positive values of $\beta$, for $w=-0.5, -1.5$, the change is nearly negligible (least being for $w=-1.5$), for $w=1$, the PBH mass will always be decreasing with time for any value of $\beta$. Furthermore, we see for $w=1$, the PBH mass decreases at a higher rate for earlier times, attaining a nearly constant value as we go towards the present time.  Thus, we see the dynamics is highly dependent not only on the model parameter $\beta$ but also on the eqaution of state parameter $w$.
\subsubsection{Modified Chaplygin Gas Model}

The modified Chaplygin gas model, an extension of the original Chaplygin gas model, has emerged as a valuable tool in cosmology and astrophysics, including the study of primordial black hole (PBH) formation and evolution. Unlike the standard Chaplygin gas, which assumes a specific form for the equation of state, the modified Chaplygin gas introduces additional parameters $A$ and $\alpha$ that allows a greater flexibility in modeling various cosmological scenarios. In the context of PBH formation, the modified Chaplygin gas model gives insights into the evolution of PBH mass by accounting for non-trivial interactions between dark energy and dark matter components.\\
 Thus, the modified Chaplygin gas model provides a framework for investigating the PBH mass evolution processes. So, we can reconstruct the equation of state of Modified Chaplygin gas. For this purpose the equation of state can be taken as,
\begin{equation}\label{23}
P=A_1\rho -\frac{C}{\rho^{\gamma}}
\end{equation}
where $A_1$, $C$ and $\gamma$ are constant parameters. When $C=0$ we recover the equation of perfect fluid, i.e. $P=A_1\rho.$ For $A_1=0$ it reduces to the generalized Chaplygin gas model. The magnitude of these parameters can provide different cosmological phases, their evolution. Here we aim to discuss the phantom phase with parametric analysis in this model. For that we consider $A_1<0$ and $C>0$. So, from Eq.(\ref{10}) and by using Eq. (\ref{18}), for modified chaplygin gas model, the evolution of PBH with respect to redshift parameter z is given by,
\begin{equation}\label{24}
\begin{split}
\frac{dM}{dz}=-\frac{1}{\alpha (1+z)\left[\frac{\beta +1}{\alpha}ln(1+z)\right]^{\frac{\beta}{\beta +1}}}\left[-\frac{a_{H}}{256\pi^{3}G^{2}M^{2}} \right.
\\
\left.
+4\pi AM^{2}\left[\rho_{0}\left(\frac{1}{1+z}\right)^{-3(1+w)} +A_1\rho_{0}\left(\frac{1}{1+z}\right)^{-3(1+w)} \right. \right.
\\
\left. \left.
-\frac{C}{(\rho_{0}(\frac{1}{1+z})^{-3(1+w)})^{\gamma}}\right] \right]
\end{split}
\end{equation}
\begin{figure}
\centering
    \subfigure[]{\includegraphics[width=0.88\linewidth]{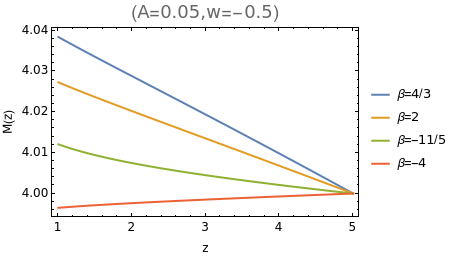}}\\
    \subfigure[]{\includegraphics[width=0.88\linewidth]{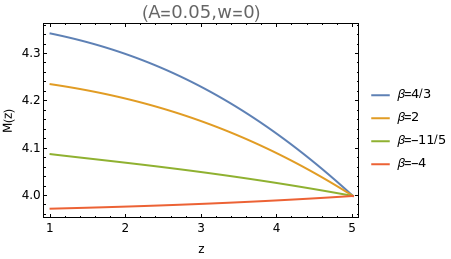}}\\
    \subfigure[]{\includegraphics[width=0.88\linewidth]{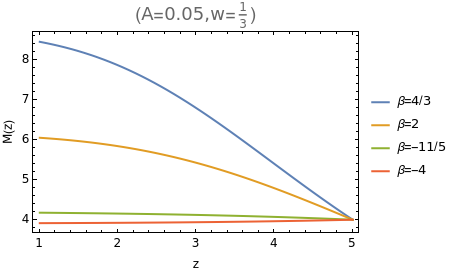}}\\
    \subfigure[]{\includegraphics[width=0.9\linewidth]{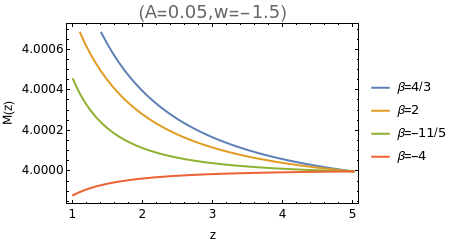}}

\caption{\it{Plot of $M(z)$ vs $z$ for Modified Chaplygin gas model for different values of $\beta$ while accretion efficiency $A$ remains fixed. For all the plots, the value of $\alpha=0.5$, $A_1=-10^{-5}, C=10^{-8}, \gamma=-0.7$ }}
\label{modi.chap1}
\end{figure}

\begin{figure}
\subfigure[]{\includegraphics[width=0.88\linewidth]{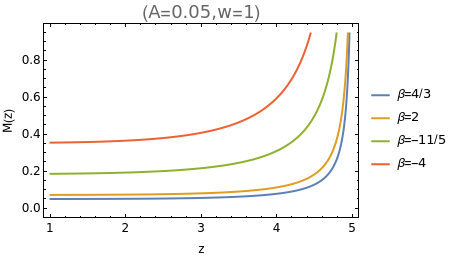}}
\caption{\it{Plot of $M(z)$ vs $z$ for Modified Chaplygin gas model for different values of $\beta$ while accretion efficiency $A$ remains fixed. For this plot, the value of $\alpha=0.5$, $A_1=-10^{1}, C=10^{-3}, \gamma=-0.7$ }}
\label{modi.chap2}
\end{figure}

In Fig. \ref{modi.chap1} and Fig. \ref{modi.chap2} we have plotted the variation of $M$ with redshift for different values of $\beta$ for a given accretion efficiency. Similar to the Chaplygin gas model, we have shown how the effective mass change of PBH changes with different values of model parameter and different values of $w$. As we can see, upon comparing with the Chaplygin gas model, the behaviour of the rate of change of PBH mass with redshift w.r.t the change in model parameter $\beta$ and $w$ follows the same pattern as the Chaplygin gas case. The conclusions thus follow.
\subsubsection{Van-Der-Waals Fluid model}

This model can discuss the evolution of PBH with just three parameters. The van der Waals fluid model, initially proposed to describe the behavior of real gases, has found applications beyond its original scope, including in the study of Black Holes. In the context of primordial black hole (PBH) formation, the van der Waals fluid model plays a crucial role in understanding the evolution of PBH mass. This model considers interactions among particles within a fluid, accounting for both attractive forces (modeled by the van der Waals term) and repulsive forces (modeled by excluded volume effects). Such interactions become significant in the early Universe when high densities and temperatures prevail. As the Universe cools and expands, critically dense regions can undergo gravitational collapse, potentially leading to the formation of PBHs. The van der Waals fluid model provides insights into the conditions under which these collapses occur, influencing the mass distribution of PBHs and their subsequent evolution.\\
So, the reconstruction of Van-Der-Waals fluid model can be done by assuming the equation of state,
\begin{equation}\label{25}
P=A_1\rho +(A_1\beta_1-\gamma_1)\rho^2
\end{equation}
This model can describe the cosmic phases and their evolution with the help of three parameters. As we aim to discuss the phantom phase, we found some constraints on the parameters\cite{sadhukhan2023non} which can be described as follows, where $A_1<0$, $\beta_1>0$ and $\gamma_1>0$. So, from Eq. (\ref{10}) and Eq. (\ref{18}), for Van-Der-Waals Fluid model, the evolution of PBH with respect to redshift parameter z is given by,
\begin{equation}\label{26}
\begin{split}
\frac{dM}{dz}=-\frac{1}{\alpha (1+z)\left[\frac{\beta +1}{\alpha}ln(1+z)\right]^{\frac{\beta}{\beta +1}}}\left[-\frac{a_{H}}{256\pi^{3}G^{2}M^{2}} \right.
\\
\left.
+4\pi AM^{2}\left[\rho_0 \left(\frac{1}{1 + z}\right)^{-3(1+w)} +A_1\rho_{0}\left(\frac{1}{1+z}\right)^{-3(1+w)} \right. \right.
\\
\left. \left.
+ (A_1\beta_1 - \gamma_1) \left(\rho_0 \left(\frac{1}{1 + z}\right)^{-3(1+w)}\right)^2\right] \right]
\end{split}
\end{equation}
\begin{figure}
\centering
    \subfigure[]{\includegraphics[width=0.88\linewidth]{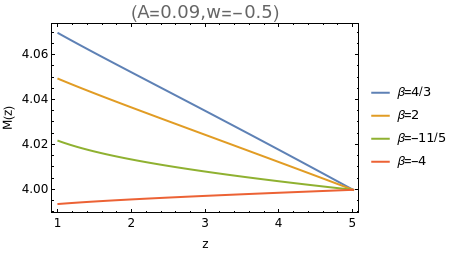}}\\
    \subfigure[]{\includegraphics[width=0.88\linewidth]{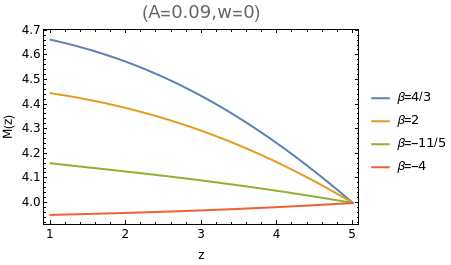}}\\
    \subfigure[]{\includegraphics[width=0.88\linewidth]{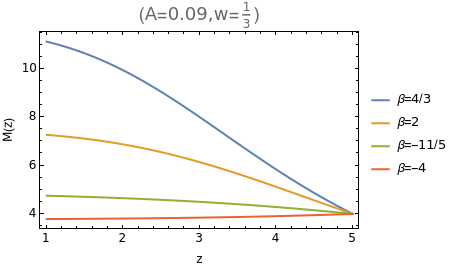}}\\
    \subfigure[]{\includegraphics[width=0.9\linewidth]{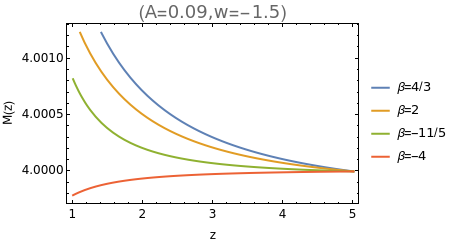}}

\caption{\it{Plot of $M(z)$ vs $z$ for van-der waals fluid model for different values of $\beta$ while accretion efficiency $A$ remains fixed. For all the plots, the value of $\alpha=0.5$, $A_1=-10^{-5}$, $\beta_1=10^{-5}$, $\gamma_1=10^{-5}$}}
\label{van.waals1}
\end{figure}

\begin{figure}
\subfigure[]{\includegraphics[width=0.88\linewidth]{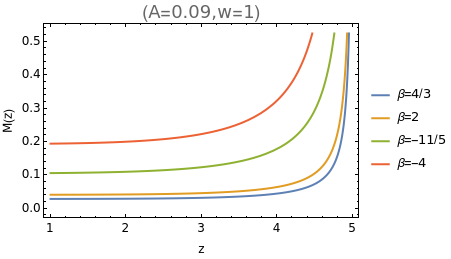}}
\caption{\it{Plot of $M(z)$ vs $z$ for van-der waals fluid model for different values of $\beta$ while accretion efficiency $A$ remains fixed. For this plot, the value of $\alpha=0.5$, $A_1=-10^{1}$, $\beta_1=10^{-5}$, $\gamma_1=10^{-5}$}}
\label{van.waals2}
\end{figure}

In Fig. \ref{van.waals1} Fig. \ref{van.waals2} we have plotted the variation of $M$ with redshift for different values of $\beta$ for a given accretion efficiency. We have shown how the effective mass change of PBH changes with different values of model parameter. We have further plotted separate graphs for different values of $w$. As we can see, irrespective of the value of $w$, for certain negative values of $\beta$ (as we have see in all the earlier cases), the PBH mass decreases with time in the presence of Vander-waals fluid like cosmic fluid. While for $w=-0.5,-1.5,0,1/3$, we can get an increase in the mass of PBH with time for positive values of $\beta$, for $w=1$, the PBH mass will always be decreasing with time for any value of $\beta$. One can also see that the rate of increase of PBH mass is much more when $w=0,1/3$ and it's very slow or nearly constant for $w=-0.5$ and $w=-1.5$ respectively.
\subsubsection{Polytropic Fluid model}

The polytropic fluid model, widely used in astrophysics and cosmology, offers a framework for understanding the behavior of various astrophysical systems, including the evolution of primordial black hole (PBH) mass. In this model, the equation of state relates the pressure of the fluid to its density by the relation given bellow, allows a wide range of behaviors depending on the value of the parameters. In the context of PBH formation, the polytropic fluid model provides valuable insights into the conditions under which gravitational collapse occurs, leading to PBH formation. By adjusting the parameters, we can explore different scenarios of PBH formation. Thus, by employing the polytropic fluid model, we can investigate the dynamics of PBH formation and their implications for cosmology and astrophysics.\\
For the reconstruction of polytropic fluid model, the equation of state should become,
\begin{equation}\label{27}
P=(A_1\beta_1-\gamma_1)\rho^2
\end{equation}
By choosing the following constraints on the constants, we will be able to discuss the phantom phase\cite{sadhukhan2023non}.
Where $A_1<0$, $\beta_1>0$ and $\gamma_1>0$. So, from Eq.(\ref{10}) and Eq. (\ref{18}), for polytropic fluid model, the evolution of PBH with respect to redshift parameter z is given by,
\begin{equation}\label{28}
\begin{split}
\frac{dM}{dz}=-\frac{1}{\alpha (1+z)\left[\frac{\beta +1}{\alpha}ln(1+z)\right]^{\frac{\beta}{\beta +1}}}\left[-\frac{a_{H}}{256\pi^{3}G^{2}M^{2}} \right.
\\
\left.
+4\pi AM^{2}\left[\rho_0 \left(\frac{1}{1 + z}\right)^{-3(1+w)} \right. \right.
\\
\left. \left.
+ (A_1\beta_1 - \gamma_1) \left(\rho_0 \left(\frac{1}{1 + z}\right)^{-3(1+w)}\right)^2\right] \right]
\end{split}
\end{equation}
\begin{figure}
\centering
    \subfigure[]{\includegraphics[width=0.88\linewidth]{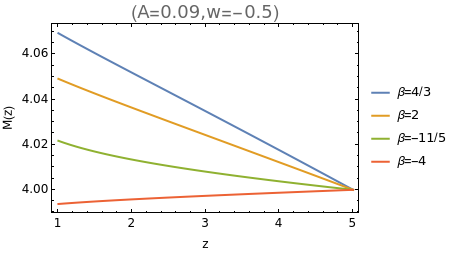}}\\
    
    \subfigure[]{\includegraphics[width=0.88\linewidth]{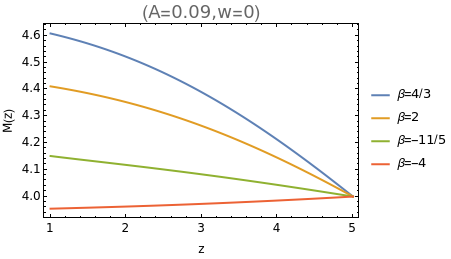}}\\
    
    \subfigure[]{\includegraphics[width=0.88\linewidth]{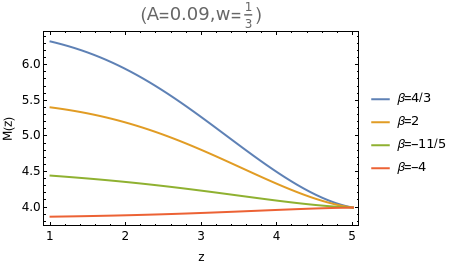}}\\
    
    \subfigure[]{\includegraphics[width=0.88\linewidth]{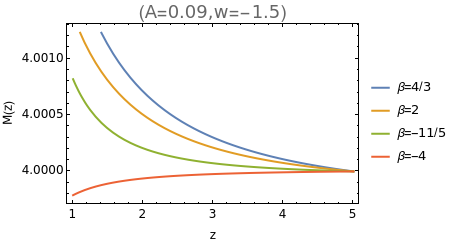}}

\caption{\it{Plot of $M(z)$ vs $z$ for polytropic fluid for different values of $\beta$ while accretion efficiency $A$ remains fixed. For all the plots, the value of $\alpha=0.5$, $A_1=-10^{5}$, $\beta_1=10^{-5}$, $\gamma_1=10^{-5}$ }}
\label{polytropic1}
\end{figure}

\begin{figure}
\subfigure[]{\includegraphics[width=0.88\linewidth]{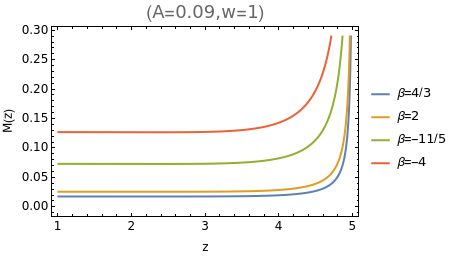}}
\caption{\it{Plot of $M(z)$ vs $z$ for polytropic fluid for different values of $\beta$ while accretion efficiency $A$ remains fixed. For this plot, the value of $\alpha=0.5$, $A_1=-10^{5}$, $\beta_1=10^{-5}$, $\gamma_1=10^{-5}$ }}
\label{polytropic2}
\end{figure}

In Fig. \ref{polytropic1} and Fig. \ref{polytropic2} we have plotted the variation of $M$ with redshift for different values of $\beta$ for a given accretion efficiency. We have shown how the effective mass change of PBH changes with different values of model parameter. We have further plotted separate graphs for different values of $w$. as we can see, irrespective of the value of $w$, for certain negative values of $\beta$ (as we have see in all the earlier cases), the PBH mass decreases with time in the presence of polytropic fluid like cosmic fluid. While for $w=-0.5,-1.5,0,1/3$, we can get an increase in the mass of PBH with time for positive values of $\beta$, for $w=1$, the PBH mass will always be decreasing with time for any value of $\beta$. One can also see that the rate of increase of PBH mass is much more when $w=0,1/3$ and its very slow or nearly constant for $w=-0.5$ and $w=-1.5$ respectively.

\section{Conclusions}
\label{Conclusions}
In this paper we investigate the effect of accretion of cosmic fluids on the evolution of primordial black holes within the framework of modified gravity theories. We consider a general form of the Hubble parameter that represents a general class of modified gravity theories and bouncing models. Our main model parameter is defined by the power $\beta$ that represents the time dependence of the Hubble parameter. We investigated the effect of the presence of different cosmological era like radiation, matter, dark energy-dominated phases like phantom and quintessence. We found that in the presence of radiation or matter or quintessence, the mass of PBH begin to evolve from a given mass for positive values of $\beta$, leading to longer life span of the PBHs and making them prone to detection.
The rate of mass accretion overpowers the loss of mass due to Hawking radiation. However, for negative values of $\beta$, we found that the PBH mass either remains constant or decreases with time, thus indicating that the presence of such background dynamics will result in decrease of the life span of the PBH. 
However, in the presence of phantom fluid, the change in mass of the PBH remains either nearly constant (for positive $\beta$) or decreases
with time (for negative $\beta$).
Since Hawking radiation and phantom energy accretion both contribute to a decrease in the mass of the PBH,  the PBH that would be expected to decay now due to the Hawking process would decay earlier due to the inclusion of the phantom energy.

We then repeated the analysis for dark energy fluids with non linear equation of state. The non linear models we considered are the Chaplygin gas, Modified Chaplygin gas, Van der Waals and Polytropic fluid. We studied the variation of mass of PBH with redshift with the model parameter $\beta$ as well as for different equation of state parameter $w = -0.5, 0, 1/3, 1, -1.5$. We found that for all these cases,
irrespective of the value of $w$, for certain
negative values of $\beta$, the PBH mass decreases with time. For 
$w = 0, 1/3$, we can get a significant increase in the mass of PBH with time for positive values of $\beta$. For $w = -0.5, -1.5$, the change is nearly negligible (least being for $w = -1.5$). And for $w = 1$, the PBH mass will always be decreasing with time irrespective of the value of $\beta$.\\
Furthermore, we found that for $w = 1$, the PBH mass decreases
initially, attaining a nearly
constant value as we go towards the present time.
\begin{acknowledgments} 
SB and AP 
would like to thank the department of physics, IIT(ISM) Dhanbad for their support during the completion of this project.
\end{acknowledgments}

\bibliographystyle{plain}
\bibliography{bibliography}

\end{document}